\documentclass[runningheads]{llncs}
\usepackage[T1]{fontenc}
\usepackage{multirow}
\usepackage{graphicx}
\usepackage{listings}
\usepackage{changepage}
\usepackage{color}
\begin{document}
\title{Llama-based source code vulnerability detection: Prompt engineering vs Fine tuning}
\titlerunning{Llama-based source code vulnerability detection}
\author{Dyna Soumhane Ouchebara\inst{1}\orcidID{0009-0002-7136-3178} \and
Stéphane Dupont\inst{1}\orcidID{0000-0003-3674-6747}}
\authorrunning{D. Ouchebara and S. Dupont}
\institute{University of Mons, Mons, Belgium
\\ \email{DynaSoumhane.OUCHEBARA@umons.ac.be} \\
\email{Stephane.DUPONT@umons.ac.be}}
\maketitle              % typeset the header of the contribution
\begin{abstract}
The significant increase in software production, driven by the acceleration of development cycles over the past two decades, has led to a steady rise in software vulnerabilities, as shown by statistics published yearly by the CVE program. The automation of the source code vulnerability detection (CVD) process has thus become essential, and several methods have been proposed ranging from the well established program analysis techniques to the more recent AI-based methods. Our research investigates Large Language Models (LLMs), which are considered among the most performant AI models to date, for the CVD task. The objective is to study their performance and apply different state-of-the-art techniques to enhance their effectiveness for this task.
We explore various fine-tuning and prompt engineering settings. We particularly suggest one novel approach for fine-tuning LLMs which we call Double Fine-tuning, and also test the understudied Test-Time fine-tuning approach. We leverage the recent open-source Llama-3.1 8B, with source code samples extracted from BigVul and PrimeVul datasets. Our conclusions highlight the importance of fine-tuning to resolve the task, the performance of Double tuning, as well as the potential of Llama models for CVD. Though prompting proved ineffective, Retrieval augmented generation (RAG) performed relatively well as an example selection technique. Overall, some of our research questions have been answered, and many are still on hold, which leaves us many future work perspectives. Code repository is available here: \url{https://github.com/DynaSoumhaneOuchebara/Llama-based-vulnerability-detection}.

\end{abstract}

\title{Llama-based source code vulnerability detection: Prompt engineering vs Fine tuning}

\section{Introduction}

While building functional software is already complex, ensuring its security is even more challenging. The push for automation and rapid development processes, enabled by the wide adoption of open-source libraries, has significantly increased software production. However, these open-source components often contain flaws, which can propagate to thousands of dependent projects. Among the most critical defects are \textit{Software Security Vulnerabilities}, which refer to faults caused by mistakes in design, development or configuration of a software system, which can be exploited by attackers to breach system security \cite{liang2025_survey}.

The number of such vulnerabilities is rising rapidly, as shown by the Common Vulnerabilities and Exposures (CVE) \cite{cve} reports in Figure \ref{fig:cve_records}. This highlights the urgent need for robust vulnerability management, and vulnerability detection (CVD) is the first crucial step in this process. Approaches have evolved from manual expert analysis to automated program analysis techniques, and more recently, to AI-based methods. Machine and Deep Learning models are indeed increasingly favored due to their ability to extract meaningful patterns from raw data, which makes them particularly interesting for vulnerability detection. The latest advances in this field involve \textit{Large Language Models (LLMs)}, which have shown exceptional performance in both natural language and software engineering tasks. Their strong reasoning and code comprehension abilities have led to promising results in recent studies applying LLMs to automated CVD.

\begin{figure}[!htbp]
\centering
\includegraphics[width=\textwidth]{images/cve_records.png}
\caption{Evolution of the number of CVEs (Common vulnerabilities and exposures) recorded from 1999 to 2024 \cite{cve_report}.}
\label{fig:cve_records}
\end{figure}

Our research is part of this broader effort, and focuses on investigating the capabilities of LLMs for the task of vulnerability detection in source code and proposing improvements, adaptations, and the use of various learning techniques to enhance their effectiveness for this particular task. Our main contributions are:
\begin{itemize}
    \item Conducting an experimental protocol where we evaluate the recent open-source Llama-3.1 8B on the vulnerability detection task on two real-world source code datasets : BigVul and PrimeVul.
    \item Investigating various Prompt engineering and Fine-tuning approaches, including Zero-shot prompting, Few-shot prompting with three approaches for example selection (random, same vulnerabilty type and RAG) and Efficient fine-tuning with QLoRA (Quantized Low Rank Adapaters). 
    \item Testing one understudied technique known as Test-Time fine-tuning, and suggesting a novel fine-tuning approach which we call Double Fine-tuning.
    \item Comparing two fine-tuning fashions for the binary classification task of vulnerability detection, namely the generative fashion and the classification fashion (explained later in the approach section).
\end{itemize}

The remainder of this paper is organized as follows: section 2 covers the state-of-the-art in LLMs and vulnerability detection in source code, section 3 presents the ideas proposed and experiments conducted, section 4 presents the results obtained along with a discussion, and section 5 concludes the paper with the answers to our research questions as well as future work perspectives.

\section{Related work}
In his section, we present a brief literature review on LLMs and vulnerability detection in source code.

\subsection{Large language models}

Large Language Models (LLMs) are the result of decades of research in language modeling, evolving through three main waves: Statistical, Neural, and Pre-trained language models \cite{llmeval2024_paper}. 
Statistical Language Models (SLMs) see text as a sequence of words and estimate its probability by computing the product of individual word probabilities, but they struggle to fully capture the richness and variability of natural language due to data sparsity \cite{minaee2024_survey}. Neural Language Models (NLMs) address data sparsity by mapping words to low-dimensional continuous vectors (embedding vectors) and neural networks. Early NLMs, such as RNNs, LSTMs, and GRUs improved NLP applications but were task-specific and could only deal with short sequences. Pre-trained Language Models (PLMs) came to adress these shortcomes and introduced the Transformer architecture which allows for parallelized processing of the sequence, consequently enabling large-scale training on vast datasets for general tasks, which we call Pre-training \cite{minaee2024_survey}. Researchers discovered that the bigger PLMs get, the more powerful they become on general-purpose tasks, and this gets us to the powerful LLMs that became the new AI standard since 2022, and which mainly refer to transformer-based PLMs that contain tens to hundreds of billions of parameters \cite{minaee2024_survey,llmeval2024_paper}. In order to allow general-purpose LLMs to adapt to some specific task in hand, two categories of techniques are available: Prompt engineering and Fine-tuning. Prompt engineering is a rapidly evolving discipline which consists of crafting the optimal input (prompt) to achieve a specific goal with a generative model \cite{minaee2024_survey}. Sometimes, when the task is too complex or very specific to particular data, prompting becomes insufficient and we need to fine-tune the model on the specific task and data in hand. But since LLMs are particularly large-sized, we must use resource efficient strategies.

\subsection{Software vulnerability detection}

Early efforts in source code vulnerability detection (CVD) were manual, relying on expert review, which, despite its accuracy, was unscalable \cite{harzevili2023_survey}. This led to the development of automated tools using static and dynamic program analysis. Static techniques \cite{flawfinder,cppcheck,sonarqube} inspect code without execution, while dynamic methods \cite{valgrind,adressSanitizer,afl} analyze runtime behavior \cite{harzevili2023_survey}. However, these approaches struggled with false positives and scalability.

To address these issues, AI-based CVD emerged around 2007 \cite{svm2007_paper}, starting with Machine Learning (ML) methods which use manually engineered features such as lexical statistics and code metrics to learn patterns from past vulnerabilities \cite{harzevili2023_survey}. While promising, these approaches required tedious feature engineering. Deep Learning (DL) alleviated this by learning representations directly from code. Models like VulDeePecker \cite{vuldeepecker2018_paper} for instance treated code as token sequences and applied RNNs. Given code’s structured nature, researchers later shifted toward graph-based representations such as Abstract Syntax Trees (ASTs) and Code Property Graphs (CPGs), processed using Graph Neural Networks (GNNs). Introduced around 2019 \cite{jiang2025_study} for CVD, GNN-based models like Devign \cite{devign2019_paper} and Reveal \cite{reveal2021_paper} achieved state-of-the-art results. More recently, the rise of Transformers and large-scale pre-trained models brought renewed interest in sequence-based modeling. Medium-sized models like CodeBERT \cite{codebert2020_paper} have been used in top-performing systems such as Linevul \cite{LineVul2022_paper} and VulBERTa \cite{hanif2022_paper}, and UniXcoder \cite{unixocder2022_paper} tested in works like \cite{ding2024_paper,xia2024_paper}. 

Current research focuses on leveraging Large Language Models (LLMs). One category of studies investigates Prompt engineering techniques. We cite a few representative works: \cite{zhou2024_survey} propose to design different prompting templates to query the close-sourced GPT-3.5 and GPT-4; \cite{nong2024_study} propose to study different techniques (zero-shot, few-shot, CoT) to query two open source LLMs including Llama-2 and Falcon and closed source ChatGPT using SARD and CVE datasets; \cite{gao2023_study} evaluate 16 LLMs using few-shot prompting for both binary and multi-class vulnerability detection on a dataset constructed from "Capture-the-flag" (CTF) challenges; \cite{steenhoek2025_study} evaluate 14 LLMs (from which Llama, Bigcode, Mistral, DeepSeek, GPT and Gemini) on SVEN dataset using different techniques (zero-shot, few-shot, CoT, contrastive in-context); \cite{feng2025_paper} propose Graph-enhanced Soft prompt tuning on CodeLlama and CodeGemma using Diversevul dataset; \cite{liu2023_paper} propose a RAG framework for GPT-3.5 using efficient retrieval techniques such as BM-25 and TF-IDF. Another category of studies investigates the performance of Fine-tuning LLMs for the CVD task. Some representative works are: \cite{du2024_paper} propose VulLLM, in which they fine-tune the open source LLMs codellama and starcoder on SVEN dataset; \cite{sultana2024_paper} evaluate fine-tuned Llama, CodeLlama, Gemma and CodeGemma on DiverseVul dataset; \cite{jiang2025_study} investigate fine-tuning Llama2, Llama3, Llama3.1 and CodeLlama on multiple CVD datasets; \cite{zhou2024_survey} experiment fine-tuning CodeLlama and Mistral on their own proposed dataset; \cite{steenhoek2025_study} evaluate fine-tuning different LLMs from Llama family, Bigcode family, and DeepSeek.

Unlike most previous research, we conduct our experiments on one LLM (Llama-3.1) using two datasets and focus on investigating the difference between various prompt engineering and fine-tuning approaches, and between the two fine-tuning fashions available with most LLMs. We also suggest one understudied technique (Test-Time tuning) and one novel approach (Double tuning) for fine-tuning LLMs. Moreover, we underline the importance of understanding each evaluation metric rather than just observing the global F1-score, and highlight the necessity of further studying the explainability of the model predictions (as part of our most urgent future work).

\section{Proposed approach}

\subsection{Problem formulation}
Code vulnerability detection is typically framed as a binary classification problem: 
\begin{math}
  X_{i} \rightarrow y_{i}
\end{math}.
Specifically, given an input source code function 
\begin{math}
    X_{i}
\end{math}, a model (neural network) predicts whether the input function is vulnerable (\begin{math}
    y_{i}=1
\end{math})
or non-vulnerable (\begin{math}
    y_{i}=0
\end{math}).
Of course, presenting the vulnerability detection problem as a binary classification task is one first step to solving the actual "real-life" problem, where we do not only want to know if the code is vulnerable, but also know the exact type of vulnerability we are facing. The problem will thus be later extended into a multi-class classification task, where the classes represent the different possible vulnerability types, generally noted by CWE (common weakness enumeration) \cite{cwe} types  in literature and available datasets.

\subsection{Datasets}
To conduct our experiments, we chose \textbf{BigVul} \cite{Bigvul_2020} and \textbf{PrimeVul} \cite{ding2024_paper} datasets. 

We justify the choice of BigVul by the fact that it is constructed from real source code projects from Github, as well as being a very well-known dataset used by most prior research, particularly state-of-art solutions. This facilitates the process of comparing our experiments with those conducted by other researchers. BigVul was created in 2020 by crawling the entries from the CVE \cite{cve} database, and linking vulnerability descriptions to publicly available GitHub repositories \cite{Bigvul_2020}. It contains 3,754 code vulnerabilities (distinct CVEs) spanning 91 different vulnerability types (CWE types) which were extracted from 348 projects mainly written in C/C++ \cite{Bigvul_2020}. Overall, the dataset, contains a total of  of 188,636 C/C++ functions with a ratio 5.7\% vulnerable and 94.3\% non-vulnerable \cite{LineVul2022_paper}. As for our experiments, we did not use BigVul dataset as-is, but applied some pre-processing as follows:
\begin{enumerate}
    \item We first extracted the columns needed by our models to function, which are only two columns: \textit{func-before} which contains the source code as text, and \textit{vul} which contains the label (0 for non-vulnerable or 1 for vulnerable).
    \item Then we split the dataset into 90\% training, 5\% validation and 5\% testing sets. Our data split is available here\footnote{\url{https://huggingface.co/datasets/DynaOuchebara/BigVul\_2columns}}.
    \item Finally, we proceeded to balancing the dataset. Though we acknowledge the limitations of using artificially balanced datasets, which do not reflect the real-world imbalance between benign and vulnerable code, this step is important to ensure our models do not trivially default to predicting the majority class. For this, we used the random under-sampling method, consisting of randomly removing instances from the majority class to match the size of the minority class. This is of course one method among others to deal with class imbalance (data augmentation, k-fold training, focal loss, etc.).
\end{enumerate}

Despite being a well-known and very utilized dataset, BigVul is 5 years old and some recent studies \cite{diversevul2023_paper,ding2024_paper} have questioned the accuracy of its labels. So to further enrich our study and confirm the confidence in our results, we reconducted all experiments on PrimeVul, which is a more recent dataset created in 2024 by merging security-related commits from many prior datasets (BigVul \cite{Bigvul_2020}, CrossVul \cite{crossvul_paper}, CVEfixes \cite{cvefixes_paper}, and DiverseVul \cite{diversevul2023_paper}) while ensuring better label accuracy with new labeling techniques, as well as reducing the possibility of data duplication. PrimeVul contains 6,968 vulnerable and 228,800 benign functions covering 140 CWEs. For our experiments, we applied to PrimeVul the same process previously described for BigVul, except for the data splitting where we used the original data split\footnote{\url{https://huggingface.co/datasets/colin/PrimeVul}} published by the authors.

\subsection{Baselines}
We chose CodeBERT \cite{codebert2020_paper} and UniXcoder \cite{unixocder2022_paper} models as baselines, which are considered as state-of-the-art models, as shown by multiple comparative studies \cite{jiang2025_study}, \cite{DeepDFA2024_paper} as well as the papers which first introduced these models for CVD (LineVul \cite{LineVul2022_paper} and SvulD \cite{SvulD2023_paper}). These models are medium-size language models with 125 million parameters. They are based on the transformer architecture, and were pre-trained on big source code corpuses. 
% Table \ref{tab:codebert_unixcoder} gives more detailed information about the size of the models, their architecture, and the data and objectives used for pre-training them.

% \begin{table*}
% \caption{Summary of important information on CodeBERT and UniXcoder models \cite{codebert2020_paper, unixocder2022_paper}.}
% \centering
% \begin{tabular}{l l l p{2.5cm} p{4cm} p{5cm}}
%     \toprule
%     Model & Params & Size & Architecture & Pre-training Objectives & Pre-training Data \\
%     \midrule
%    CodeBERT & 125M & $\sim$500MB & 
%     Transformer (RoBERTa-base) Encoder-only &  
%     Masked Language Modeling, Replaced Token Detection & 
%     2.1M bimodal (Natural Language-Code pairs) and 6.4M unimodal code samples from CodeSearchNet dataset covering Python, Java, JavaScript, PHP, Ruby, Go. \\

%     \rule{0pt}{15pt} UniXcoder & 125M & $\sim$500MB & 
%     Transformer (RoBERTa-base) Encoder-Decoder &
%     Masked Language Modeling, Unidirectional Language Modeling, Denoising Auto-Encoding, Multi-modal Contrastive Learning, Cross-Modal Generation & 
%     2.3M functions from CodeSearchNet (Natural Language-Code pairs) in Ruby, Java, Python, PHP, Go, and JavaScript. AST representations and code comments were included. \\
%     \bottomrule
% \end{tabular}
% \label{tab:codebert_unixcoder}
% \end{table*}

To prepare our baselines, we decided not to take results from existing papers who have tested these models before us, due to the lack of unification observed in these papers (almost every paper presents different results due to the different experimental setup and parameters). We finetuned them on BigVul (resp. PrimeVul) training set, and then tested the fine-tuned models on the test set. 

\subsection{Approach} \label{section:experiments}

As described earlier in the introduction section, the motivation behind studying the potential of LLMs for vulnerability detection lies in the fact that these models are first of all pre-trained on vast amounts of textual data, among which natural language and programming code corpuses. Consequently, there is a high probability that these corpuses contain data related to security issues such as source code vulnerabilities. This prior knowledge makes LLMs an interesting starting point for building a CVD solution. Llama \cite{llama} models, in particular, are a good choice for they are open-source and efficient. In fact, unlike proprietary models, they can be fine-tuned on security-specific datasets, allowing for an improved accuracy in tasks like vulnerability detection. Llama models are also optimized for inference, making them cheaper to deploy compared to larger models like GPT-4 or Claude.
They provide a good balance between model size, latency and accuracy. From the Llama series, we chose to conduct our experiments on the Llama-3.1 8B version from Llama 3 series \cite{llama32024_paper}. The 3.1 version is the latest version of Llama available in Europe which proposes a "medium" sized LLM like the 8B one, the 3.2 version being only available in USA and Canada at the moment and the 3.3 version proposing only bigger models (over 70B). Moreover, 8 billion parameters is convenient because it is large enough to understand complex code patterns, yet small enough for cost-effective deployment and inference. The idea is to test different techniques to make Llama-3.1 8B more suitable for our CVD task. Two main categories of approaches for adapating LLMs are studied: Prompting and Fine-tuning.

\subsubsection{Prompting : Adapting LLMs without changing their weights}
Prompting is the process of guiding an LLM’s behavior by crafting well-structured inputs (prompts). Instead of modifying the model’s parameters, we use cleverly designed prompts to get the model to generate useful outputs. We explore two main prompting techniques in our study : Zero-shot and Few-shot prompting.

In Zero-Shot Prompting, we only give the instruction to the model. No examples are given and the model relies only on its pre-trained knowledge. Generally, this technique works well for general knowledge questions, but performance may be poor for specialized tasks. Our zero-shot prompt is the following: 
    
\begin{lstlisting}[language=Python, caption=Zero-shot prompt, label=lst:zero-shot, basicstyle=\small]
""" Classify the source code into Vulnerable or Safe, and return the answer as the corresponding label.
Code: #code snippet we want to predict
Label: """
    \end{lstlisting}

In Few-Shot Prompting, we give the instruction to the model in addition to a few labeled examples to guide the model. The model learns the pattern from examples and this helps it better understand the expected format of the answers it should return. Our few-shot prompt is the following: 

\begin{lstlisting}[language=Python, caption=Few-shot prompt, label=lst:zero-shot, basicstyle=\small]
""" Classify the source code into Vulnerable or Safe, and return the answer as the corresponding label. Here are some examples:
Code: #example code 1
Label: #example label 1
Code: #example code 2
Label: #example label 2
...
Code: #code snippet we want to predict
Label: """
    \end{lstlisting}

To constitute the prompt, for each test code, we choose 6 examples from the training set. We initially conducted our few-shot experiments with 4, 6 and 10 examples, but we will only report the results obtained with 6-shot prompts because it proved most performant and efficient. We followed 3 strategies to select these examples. In the first one, we randomly choose 3 vulnerable code examples and 3 safe code examples from the training set. In the second one, we choose 3 vulnerable examples that correspond to the same type of vulnerability (CWE type) of the test code, and we randomly choose the 3 safe examples. In the third one, we use Retrieval Augmented Generation (RAG) \cite{rag}. We first generate embeddings for all code snippets in the training set using an embedding model for code; we chose CodeBERT for its excellent code understanding capability. We save these embeddings in an index (using FAISS\footnote{https://faiss.ai/} library), then for each test code, we search for the 6 most similar code snippets in the training set leveraging that index, using L2 (Euclidean) distance as a similarity measure.

\subsubsection{Fine-tuning: Adapting LLMs by tuning their weights} 
Fine-tuning involves training an LLM on a custom dataset so that it adapts to a specific task, in our case CVD. Two main approaches exist: Full Fine-tuning and Efficient Fine-tuning. For our experiments, we tested Efficient Fine-tuning with LoRA as well as Quantization, because Full Fine-tuning requires too much GPU memory requirements (since it updates all the weights of a billion parameter model).

LoRA \cite{lora2021_paper} is a method that adds small, trainable adapter layers instead of modifying all model weights. This approach requires less resources than Full Fine-tuning while maintaining performance. Quantized LoRA (QLoRA) is an even more memory-efficient version of LoRA. Instead of working with the full-precision model (32-bit), we apply 4-bit quantization to the model, which reduces its memory footprint.
It is important to note that since we are using a compressed model in addition to LoRA, the performance of the resulting fine-tuned model does not match a fully fine-tuned model, but still, the performance drop is usually not too penalizing if the right hyperparameters are chosen.

We study 3 different approaches for fine-tuning. 

The first approach is the classic training then testing, where we first train the model on our whole training data (using QLoRA), and then test the fine-tuned model on the test data. To do the training phase for our binary classification task of vulnerability detection, we have two options. The first one is to fine-tune the LLM as a generative model and then analyze the textual response generated and see if it is "Vulnerable" or "Safe". The second one is to add a classification head to the model, which consists of a feed-forward neural network (FFNN) with one output neuron which returns the probability of vulnerability. We consequently tested both fine-tuning fashions. As for the approaches that follow, we applied the second fine-tuning fashion (classification head).

The second approach is Test-Time fine-tuning, where for each test sample, we retrieve 6 similar examples from the training data (using RAG just like explained for the 3rd few-shot learning strategy), then we do a quick fine-tuning of the model using only these examples. The idea is that instead of just adding the examples to the prompt and relying on the model to effectively leverage the information present in the input to generate the most suitable output, we can use the examples to actually change the model weights which have a more direct effect on the answer generated. Another benefit of this technique is that we can use as many examples as we want for the fine-tuning, whereas we are limited by the maximum context length when adding the examples to the prompt. The idea of Test-time training was studied by a few researches \cite{akyürek2025surprisingeffectivenesstesttimetraining,sun2020testtimetrainingselfsupervisiongeneralization,hübotter2025efficientlylearningtesttimeactive}.

Finally, the third approach merges the two previous ones, where we first train the model on the whole train data, then we further tune the model at test-time using the closest training samples. We call this Double fine-tuning.

\subsubsection{Models used}
We specifically experimented on two models.

Llama-3.1 8B Base \cite{llama3.1}, is a pre-trained only version of the model. The model has been pre-trained on a massive corpus of text in an unsupervised manner on next-word prediction (auto-regressive language modeling). It acts as a foundation model for further specialization (tuning) on custom datasets. 

LLama-3.1 8B Instruct \cite{llama3.1_instruct} is a fine-tuned version of the previously described base model using supervised instruction datasets. It is thus optimized for zero-shot and few-shot prompting settings. Instruct models can also be fine-tuned further to be more performant on a specific task, but with some challenges (it is important to carefully adapt it to our task without loosing its instruction-following ability, ie. catastrophic forgetting).

So in our experiments, for prompting, we used Llama-3.1 8B Instruct, and for Fine-tuning, we used both Llama-3.1 8B Base and Llama-3.1 8B Instruct. 

\section{Results and discussion}

\subsection{Evaluation metrics}
When evaluating a binary classification model for vulnerability detection, we need to carefully interpret the following key evaluation metrics.

\textbf{Accuracy} measures the overall correctness of predictions, ie. the overall proportion of correctly classified instances (both Safe and Vulnerable) out of all instances. While accuracy gives a general sense of model performance, it can be misleading if the dataset is imbalanced. Since we are working on a balanced dataset, this is not our case, however, it still does not tell us whether the model is better at detecting vulnerabilities or avoiding false alarms. In order to answer these questions, we must calculate other metrics, which follow. 

\textbf{Precision} measures how many of the instances predicted as "Vulnerable" are actually vulnerable. A high precision means that when the model says "this code is vulnerable", it is usually correct, ie. the model makes fewer false alarms.

\textbf{Recall} measures how many of the actual "Vulnerable" instances were detected. A high recall means that the model catches most vulnerabilities.

\textbf{F1-score} is the harmonic mean of "Precision" and "Recall", balancing both metrics. If both avoiding false alarms and detecting as much vulnerabilities as possible are important, F1-score is the best single metric to consider.

The \textbf{ROC curve} (Receiver Operating Characteristic Curve) is a graphical representation of a classifier’s performance across different decision thresholds. It plots the True Positive Rate (TPR) on the y-axis against the False Positive Rate (FPR) on the x-axis. A classifier with a perfect separation of classes will have a ROC curve that reaches the top-left corner (with recall close to 1 and precision close to 1), while a random classifier produces a diagonal line. The \textbf{AUC} (Area Under the Curve) is a numerical metric ranging from 0 to 1 calculated from the ROC curve, which quantifies how well a classifier separates positive and negative classes. A higher AUC indicates better model performance.

For CVD in a general context, high recall (for vulnerable class) is often desirable to catch as many vulnerabilities as possible, while keeping precision (for vulnerable class) high to reduce false alarms. Consequently, we should watch both metrics to make a good interpretation of how good each model works. F1 score being helpful to find a good trade-off between Recall and Precision, and AUC being the metric that best summarizes the classification performance of a model, we will consider these as metrics to globally compare our models.

\subsection{Experimental results and discussion}
Now that we thoroughly explained our experiments and the way we are evaluating them, we can review Table \ref{tab:results} which presents the results.

\begin{table*}[t!]
    \caption{Performance comparison between baselines and our proposed approaches.}
    \label{tab:results}
    \renewcommand{\arraystretch}{1.2}
    \setlength{\tabcolsep}{3pt}
    \centering
    \begingroup
    \fontsize{7}{8}\selectfont
    \makebox[\linewidth]{%
    \begin{tabular}{|l|l|l|c|cc|cc|ccc|}
        \hline
        Model & Technique & Data & Accuracy & \multicolumn{2}{c|}{Precision} & \multicolumn{2}{c|}{Recall} & \multicolumn{3}{c|}{F1 Score} \\
        & & & & 0 & 1 & 0 & 1 & 0 & 1 & Avg \\
        \hline
        \multicolumn{11}{l}{\textbf{Baselines}} \\
        \hline
        \multirow{2}{*}{CodeBERT} & \multirow{2}{*}{Full Fine-tuning} & BigVul & 0.920 & 0.91 & 0.93 & 0.93 & 0.91 & 0.920 & 0.920 & 0.920 \\
        \cline{3-11}
        & & PrimeVul & 0.761 & 0.73 & 0.79 & 0.81 & 0.70 & 0.770 & 0.740 & 0.760 \\
        \hline
        \multirow{2}{*}{UniXcoder} & \multirow{2}{*}{Full Fine-tuning} & BigVul & 0.943 & 0.94 & 0.94 & 0.94 & 0.94 & 0.940 & 0.940 & 0.940 \\
        \cline{3-11}
        & & PrimeVul & \textbf{0.770} & 0.77 & \textbf{0.77} & 0.77 & 0.78 & \textbf{0.770} & 0.770 & \textbf{0.770} \\
        \hline
        \multicolumn{11}{l}{\textbf{Our proposed approaches}} \\
        \hline
        \multirow{2}{*}{Llama-3.1 8B instruct} & \multirow{2}{*}{Zero shot} & BigVul & 0.399 & 0.31 & 0.43 & 0.16 & 0.64 & 0.211 & 0.514 & 0.363 \\
        \cline{3-11}
        & & PrimeVul & 0.510 & 0.55 & 0.51 & 0.11 & 0.91 & 0.180 & 0.650 & 0.410 \\
        \hline
        \multirow{2}{*}{Llama-3.1 8B instruct} & \multirow{2}{*}{Few shot random} & BigVul & 0.588 & 0.56 & 0.74 & 0.90 & 0.28 & 0.690 & 0.400 & 0.590 \\
        \cline{3-11}
        & & PrimeVul & 0.640 & 0.60 & 0.74 & 0.84 & 0.44 & 0.700 & 0.550 & 0.640 \\
        \hline
        \multirow{2}{*}{Llama-3.1 8B instruct} & \multirow{2}{*}{Few shot same CWE type} & BigVul & 0.582 & 0.55 & 0.72 & 0.90 & 0.27 & 0.680 & 0.390 & 0.540 \\
        \cline{3-11}
        & & PrimeVul & 0.648 & 0.60 & 0.75 & \textbf{0.86} & 0.44 & 0.710 & 0.560 & 0.630 \\
        \hline
        \multirow{2}{*}{Llama-3.1 8B instruct} & \multirow{2}{*}{Few shot RAG} & BigVul & 0.700 & 0.69 & 0.71 & 0.73 & 0.67 & 0.710 & 0.692 & 0.700 \\
        \cline{3-11}
        & & PrimeVul & 0.670 & 0.66 & 0.68 & 0.70 & 0.64 & 0.680 & 0.660 & 0.670 \\
        \hline
        \multirow{2}{*}{Llama-3.1 8B instruct} & \multirow{2}{*}{RAG + Test-Time fine-tuning} & BigVul & 0.780 & 0.82 & 0.75 & 0.72 & 0.84 & 0.770 & 0.792 & 0.780 \\
        \cline{3-11}
        & & PrimeVul & 0.690 & 0.74 & 0.67 & 0.61 & 0.78 & 0.670 & 0.720 & 0.690 \\
        \hline
        \multirow{2}{*}{Llama-3.1 8B instruct} & \multirow{2}{*}{Efficient Fine-tuning with QLoRA} & BigVul & 0.900 & 0.88 & 0.94 & 0.94 & 0.86 & 0.910 & 0.900 & 0.900 \\
        \cline{3-11}
        & & PrimeVul & 0.573 & 0.57 & 0.58 & 0.62 & 0.53 & 0.590 & 0.550 & 0.570 \\
        \hline
        \multirow{2}{*}{Llama-3.1 8B base + classification head} & \multirow{2}{*}{Efficient Fine-tuning with QLoRA} & BigVul & 0.949 & 0.96 & 0.94 & 0.94 & 0.96 & 0.950 & 0.950 & 0.950 \\
        \cline{3-11}
        & & PrimeVul & 0.740 & 0.72 & 0.77 & 0.79 & 0.69 & 0.750 & 0.730 & 0.740 \\
        \hline
        \multirow{2}{*}{Llama-3.1 8B base + classification head} & \multirow{2}{*}{Double fine-tuning} & BigVul & \textbf{0.970} & \textbf{0.96} & \textbf{0.98} & \textbf{0.98} & \textbf{0.96} & \textbf{0.970} & \textbf{0.970} & \textbf{0.970} \\
        \cline{3-11}
        & & PrimeVul & 0.768 & \textbf{0.82} & 0.73 & 0.68 & \textbf{0.85} & 0.750 & \textbf{0.790} & \textbf{0.770} \\
        \hline
    \end{tabular}
    }
    \endgroup

    \vspace{3pt}

    \begin{adjustwidth}{-3cm}{-3cm}
    \footnotesize
    Double fine-tuning of Llama-base with a classification head yields the best performance of 0.97 F1-score on BigVul, exceeding the baselines. Zero-shot and Few-shot prompting are overall unsatisfactory, though RAG is relatively performant, whether it is used for Few-shot prompting or for Test-time fine-tuning, and the latter is the most performant option. Fine-tuning the base model with a classification head ("classifier fashion")  gives better results than fine-tuning the instruct model ("generative fashion"). Overall results on PrimeVul data mostly show the same behavior as on BigVul, except that the best approach (double fine-tuning) only matches UniXcoder baseline with an 0.770 average F1-score without exceeding it.
    \end{adjustwidth}
\end{table*}

As for the baselines, we observe that both \textbf{CodeBERT} and \textbf{UniXcoder} models, when fully \textbf{fine-tuned} on the BigVul training data, have an excellent ability to detect vulnerabilities reaching a performance of 0.92 F1 score for CodeBERT and 0.94 F1 score for UniXcoder for "vulnerable" class on the test data. 
% PRIMEVUL
The detection ability is reduced on PrimeVul dataset to 0.74 and 0.77 F1 score for CodeBERT and UniXcoder respectively. This is expected since the authors of the dataset have observed the same behavior over different CodeLMs. This suggests that the models cannot effectively learn from the more complex and realistic distribution of vulnerabilities in PrimeVul, which is a more challenging evaluation environment than most previous benchmarks \cite{ding2024_paper}. We note, however, that we reach a very good performance compared to the original paper (which tested other models than those we test in our research).

As for our proposed approaches, \textbf{Llama-3.1 8B instruct with zero-shot prompting} achieves a medium F1 score of 0.514 for "vulnerable" class with BigVul data. However, the low precision and recall of "safe" class reveal that this result is not due to a medium detection capability, but rather to a bias toward predicting "vulnerable" for most input codes. The same behavior is observed on PrimeVul data. This could indicate that the model has some knowledge about what vulnerabilities are (probably gained from their large pre-training on various types of text, among which text related to cybersecurity as well as code corpuses), but this knowledge is not enough to make precise predictions.
    
Zero-shot prompting results being unsatisfactory, we tested \textbf{Few-shot prompting}, suggesting that giving examples of vulnerable and safe code to the model would help make more accurate predictions. With the first strategy where we randomly sample examples, we indeed observe some improvement, however it only concerns the previous bias of the model towards predicting "vulnerable" which is no longer present. The model does not necessarily recognize vulnerable code better, but it does recognize safe code better, with an F1 score for "safe" class improved from 0.211 to 0.690 on with BigVul data. The results on PrimeVul show the same behavior. Few-shot prompting thus proved helpful. We then tested the second strategy, suggesting that having examples of the same vulnerability type could better help the model recognize the vulnerability in our test code. This hypothesis was proven wrong, as the performance did not improve when compared to random sampling on both datasets. The last strategy using RAG was the most effective one, with an enhanced average F1-score of 0.700 against 0.590 and 0.540 for the first and second strategy respectively on BigVul. We particularly note a better recall of 0.67 for "vulnerable" predictions and an overall improved recognition of "safe" code (0.710 F1-score). The same observation is made on PrimeVul. This suggests that choosing the most similar code snippets to our test code based on embeddings is a relatively good approach.

We then followed with \textbf{Test-Time (TT) Fine-tuning} using the examples retrieved by the RAG system, suggesting it would further improve the capacity of the model to benefit from the examples. The detection capacity indeed improved with an F1-score of 0.792 for "vulnerable" class against 0.692 with Few-shot RAG on BigVul, and the same observation is made on PrimeVul. This suggests that using examples to quickly train the model before inference is more effective than feeding them through the prompt.

To see if performance can be further improved, we studied more "complete" fine-tuning approaches. We first tested the first fine-tuning approach (generative fashion) on \textbf{Llama-3.1 8B instruct} using  \textbf{Efficient Fine-tuning using QLoRA}. The expectation was to get better accuracy since the model gets a supplementary training on a whole vulnerability detection dataset. The performance indeed improved significantly, from 0.780  (TT fine-tuning) to 0.900 average F1-score. This observation is however not made with PrimeVul.
We thus tested the second fine-tuning approach which consists of using \textbf{Llama-3.1 8B base and adding a classification head} which classifies the code into 0 (safe) or 1 (vulnerable). The performance improved even further, yielding the best results among all previous experiments, with an F1-score of 0.95, precision of 0.94 and recall of 0.96 for "vulnerable" class on BigVul. This performance is comparable to the fine-tuned UniXcoder, rather slightly better (0.95 versus 0.94). What we can deduce from this is that, for our vulnerability classification task, feeding the embeddings generated by the fine-tuned LLM into a binary classifier (FFNN) is more effective than using these embeddings for text generation and observing the generated text. This conclusion also applies even more to Primevul, where we observe an importantly improved average F1-score of 0.740 against 0.570.

Further fine-tuning this model at Test-time with examples retrieved by the RAG system improved the detection capacity to reach an F1-score of 0.970 on BigVul data, making this final \textbf{Double Fine-tuning} approach the most effective one. This approach also gives best results on PrimeVul data with an average F1-score of 0.770, however it only matches the best baseline UniXcoder. Conducting a root-cause analysis and assessing more datasets would help us justify this gap and make a more general conclusion as to the effectiveness of the approach. We still note that it slightly exceeds UniXcoder in terms of F1-score for "vulnerable" class which is class which interests us most, with 0.79 against 0.77.

It is important to note that all metrics reported above are based on one dataset split. To further improve the statistic rigor of our experimental evaluation, we will include confidence intervals in our future contributions.

\begin{figure*}[h]
  \begin{adjustwidth}{-2cm}{-2cm}
  \centering
  \includegraphics[width=\linewidth]{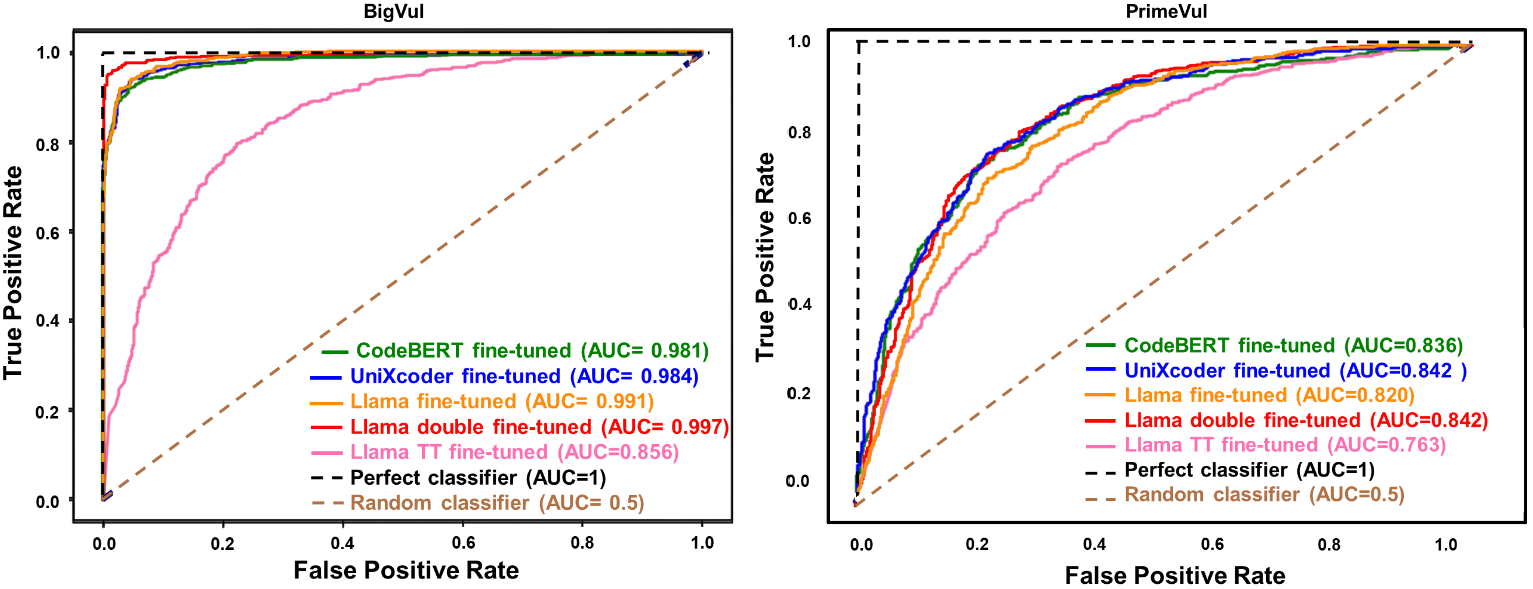}
  \caption{Comparative ROC curves between our best performing models and baselines.}
  \vspace{3pt}
  {
  \begin{adjustwidth}{+1cm}{+0cm}
  \footnotesize
  Our double fine-tuned Llama-base with a classification head yields best performance as it is closest to the perfect classifier in the upper left corner, with an AUC value of 0.997 on BigVul data. This approach also performs best on PrimeVul data, however it only matches UniXcoder baseline with an AUC of 0.842.
  \end{adjustwidth}
  \par}
  \label{fig:roc}
  \end{adjustwidth}
\end{figure*}

To further analyze our best performing approaches we represented the ROC curve for our different fine-tuning approaches as well as the fine-tuned CodeBERT and UniXcoder baselines. Figure \ref{fig:roc} shows the comparative ROC curves. With BigVul data, we observe that our double fine-tuning and one-step fine-tuning approaches with Llama have near-perfect performance, as they are close to the upper left corner, exceeding the baselines, where double fine-tuning achieves an AUC of 0.997 against 0.981 and 0.984 for CodeBERT and UniXcoder, respectively. This means that the model has an excellent capacity to discriminate safe and vulnerable classes. As for test-time fine-tuning, it is far behind the other models, which is expected since the model is only fine-tuned for the 6 closest examples instead of the whole dataset, but it is still a relatively effective model compared to prompting.
On PrimeVul, results show the same behaviour, except our best model only matches the best baseline UniXcoder (0.842 AUC). 

\begin{figure*}[h]
\begin{adjustwidth}{-4cm}{-4cm}
\centering
\includegraphics[width=\linewidth]{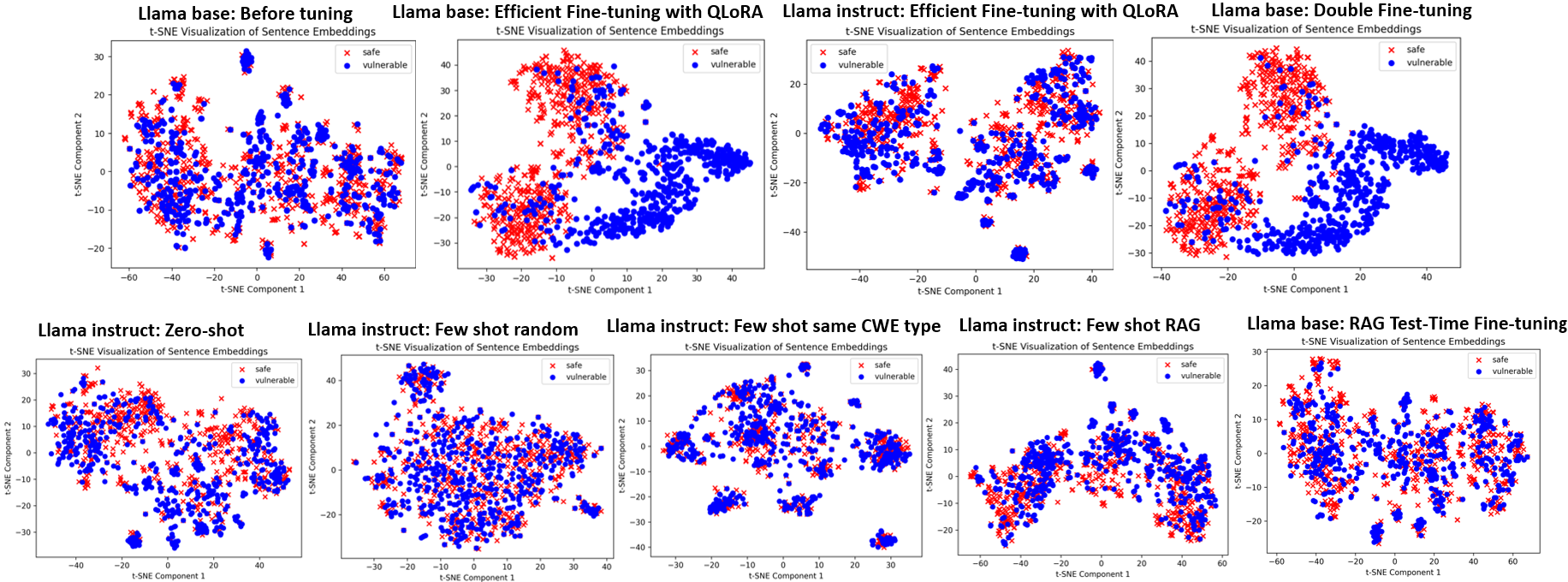}
\caption{Comparison between t-SNE plots for our proposed models on BigVul dataset.}
\vspace{3pt}
  {\raggedright
  \begin{adjustwidth}{+0.5cm}{+0.25cm}
  \footnotesize
  The only models capable of generating embeddings that are sufficiently discriminative are the double fine-tuned and one-step fine-tuned Llama-base with a classification head, where we clearly see two groups in the plot (the blue group corresponding to vulnerable samples and red group corresponding to safe samples).
  \end{adjustwidth}
  \par}
\label{fig:tsne}
\end{adjustwidth}
\end{figure*}

To confirm the impact of fine-tuning on the discrimination capacity of the model, we can represent t-SNE plots using the embeddings generated by the model before fine-tuning and after fine-tuning. These embeddings are retrieved at the last layer before the classifier layers, where we have an embedding vector of size 4096 for each token in the input sequence. We chose to use the mean of these embeddings as an embedding for the whole sequence.
Figure \ref{fig:tsne} presents the t-SNE plots for our best performing models on BigVul before and after tuning, as well as our other proposed approaches using Llama-instruct. We do not include PrimeVul plots for a lack of space, but the conclusions are fairly similar.

T-SNE is a dimensionality reduction technique which allows us to represent high dimensional embeddings in a 2D or 3D space. For a performant classifier, test samples within the same class should be represented close to each other, and samples from different classes should be far from each other. As shown by Figure \ref{fig:tsne}, the only models capable of generating embeddings that are sufficiently discriminative are the double and the one-step fine-tuned Llama-base with a classification head, where we clearly see two groups in the plot (the blue group corresponding to vulnerable samples and red group corresponding to safe samples). There are still a few samples for which the embeddings do not capture the corresponding class, mainly vulnerable samples which seem to "look like" safe ones (blue dots in the red group). Further investigation of the actual source code corresponding to these samples (failure cases) may help us understand the reason behind this behavior. Comparing the plot before fine-tuning Llama-base, where we do not see any clear separation between the classes, and after fine-tuning, implies that fine-tuning on a dataset specific to the CVD task is more essential to solve the task than possessing general code-related knowledge. 

\subsection{Takeaways}
From these results, we can keep the following takeaways.

\textbf{Fine-tuning is crucial}: Llama performs poorly in untuned, zero-shot or few-shot settings for vulnerability detection task. This is probably due to the fact that techniques like zero-shot and few-shot prompting only allow the model to retrieve information from its past knowledge, such as this information is most relevant to the task in hand. If this prior knowledge is insufficient to solve the problem, giving examples will not help in any way more than simply guiding the model output-format wise. Fine-tuning, on the other hand, allows for gaining new knowledge and this is what significantly boosts performance. So, our hypothesis is that the CVD task is complex enough to require the model to be trained specifically for this task and on data relevant to it.
Nevertheless, a useful result to retain from our different Few-shot prompting experiments is that RAG seems to be the best approach for choosing examples and yields relatively good results, in addition to the fact that Test-Time fine-tuning is an even more effective way to benefit from the examples rather than simply adding them to the prompt.
Of course, it is important to note that our conclusions are only made regarding the model and datasets we tested, we still need to experiment on other models and datasets to state this as a general conclusion about LLMs in CVD.
    
\textbf{Llama-3.1 models show strong potential}: Our results suggest that a properly fine-tuned Llama 3.1 8B can match, even exceed current state-of-the-art CVD models. Now, one may ask again why such a big sized model is interesting if its performance only slightly exceed the lighter baselines. The answer is that:
\begin{itemize}
    \item An LLM like Llama-3.1 is potentially better at reasoning over complex patterns. Unlike UniXcoder, which is strictly trained on code, Llama-3.1 has been exposed to a wide range of knowledge, which might help in detecting subtle vulnerability patterns that require contextual understanding. But this is only a hypothesis which we will later verify by testing the two models on datasets containing more subtle vulnerabilities than those present in BigVul and PrimeVul (though PrimeVul has already proven to be quite complex).
    \item An LLM like Llama-3.1 is a more flexible and versatile model which might allow us to generalize the solution to a broader security workflow in the future (e.g., vulnerability reasoning, security report generation, etc.). We can thus imagine a solution that encompasses the complete vulnerability management cycle in one tool based on one unique model, which would be more convenient for users who will not have to juggle between different softwares. But the real-life applicability of the solution will still need to be verified first (through manual auditing and CVE rediscovery experiments).
    \item An LLM-based solution leaves the door open for exploring many recent LLM techniques, which are constantly evolving at a faster pace than the techniques proposed in the range of medium sized pre-trained models.
    \item Finally, if the only point in disfavor of Llama 3.1 8B is its big size compared to UniXcoder-like models, it is important to note that there is currently a constant effort in making smaller LLMs as effective as bigger ones. For instance, Llama3.2 3B is not too far in performance from to Llama3.1 8B in benchmarks \cite{llama3.2}, despite being almost three times smaller. There are also multiple ways to "compress" a model into a smaller one while keeping most of its capabilities. One such technique is knowledge distillation, which can for instance be applied to compress Llama-3.1 8B into a Llama-3.1 1B, while keeping an important percentage of its capabilities \cite{distillation}. We, however, state that the hypothesis we make as to the potential efficiency and performance of these smaller variants of Llama needs to be concretely verified.
\end{itemize}

\section{Conclusion}

Through experiments conducted with the Llama-3.1 8B model on BigVul and PrimeVul dataset, we have demonstrated that LLMs hold significant potential for vulnerability detection, matching (on PrimeVul) rather even surpassing (on BigVul) current state-of-the-art models like CodeBERT and UniXcoder. Our findings highlight that simple prompting techniques such as zero-shot and few-shot learning are insufficient to extract meaningful CVD capabilities from LLMs like LLama-3.1 8B, reinforcing the importance of specialized training to achieve competitive performance. Efficient fine-tuning methods like QLoRA have proven to be key to optimizing performance while maintaining computational feasibility. We proposed the novel Double fine-tuning technique, which proved to be the most performant approach among all those we tested. We also observe that for our CVD task, the classifier fashion for fine-tuning (binary classifier on top of the LLM) is more effective than the generative fashion. Finally, while prompting techniques were ineffective, a useful result to retain is that RAG seems to be the best approach for example selection. We also suggested a rather unexplored technique to benefit from the selected examples, which is Test-Time fine-tuning, and which gave better results than few-shot prompting.

While we have answered a few of our initial research questions, many remain to be explored. Future work will focus on improving model interpretability through explainability techniques and failure case investigation (per-CWE analysis), analyzing the cost/scalability concerns for practical deployment of our solution (LLMs being resource-intensive) in terms of training/inference latency, GPU, carbon footprint and possible optimizations to reduce compute, experimenting with more advanced prompting techniques, and testing different datasets (other C/C++ datasets, other programming languages, more interestingly memory-safe languages) to assess the model's generalization capacity. In addition to that, more LLMs and alternative architectures, such as CodeLlama, DeepSeekCoder and emerging Graph LLMs, will be evaluated to determine the most suitable LLMs for CVD, but also to better point out the value of our proposed double fine-tuning strategy by making sure to distinguish the gains stemming from the model from those due to the strategy itself. Beyond these short-term objectives, we plan to expand our research towards multi-class vulnerability classification, and with a bigger vision, to integrating LLMs into a broader vulnerability management workflow encompassing vulnerability assessment and/or remediation phases. We also plan to broaden the security relevance dimension of our research by studying the real-world exploitability and SDLC integration of our solution. Investigating dynamic vulnerability detection methods using AI, an understudied area in literature, is also another path that we could explore.

\begin{credits}
\subsubsection{\ackname} This study was funded by CyberExcellence project of CyberWal program by Digital Wallonia.

\subsubsection{\discintname}
The authors have no competing interests to declare that are
relevant to the content of this article.
\end{credits}

\bibliographystyle{splncs04}
\bibliography{bibliography}

% Appendix
\appendix
\section{Setup and parameters}

Table \ref{tab:parameters} details the parameters used in the experiments. As for the setup, we used a system with the following resources: RTX 6000 Ada GPU with 40GB of VRAM, 100GB of RAM, 64GB of disk space.

\begin{table}
    \caption{Parameters used for training.}
    \label{tab:parameters}
    \centering
    \begin{tabular}{|l|l|}
        \hline
        Parameter & Value \\
        \hline
        \textbf{QLoRA parameters} & \\
        Quantization & 4-bit \\
        Rank (r) & 16 \\
        Scaling (alpha) & 8 \\
        Target modules & 'q\_proj', 'k\_proj', 'v\_proj', 'o\_proj' \\
        \textbf{Training parameters} & \\
        N° epochs & 4 \\
        Batch size & 16 \\
        Optimizer & paged\_adamw\_32bit \\
        Learning rate & 2e-4 \\
        \textbf{Other parameters}  & \\
        Rag retrieval depth & 6 examples \\
        \hline
    \end{tabular}
\end{table}

\end{document}